\documentstyle[onecolumn,epsfig]{mn}

\def\be{\begin{equation}}
\def\ee{\end{equation}}
\def\lmf{\overline{\log M}_f}
\def\lmi{\overline{\log M}_i}
\def\lm{\overline{\log M}}
\def\mfmi{M_{GCS,f}/M_{GCS,i}}
\def\nfni{N_{f}/N_{i}}
\def\dmio{\Delta \overline{\log M}_{inn-out}}
\def\sgf{\sigma_f}
\def\sg{\sigma}
\def\ltorder{\hbox{ \rlap{\raise 0.425ex\hbox{$<$}}\lower
0.65ex\hbox{$\sim$} }} 
\def\gtorder{\hbox{ \rlap{\raise 0.425ex\hbox{$>$}}\lower
0.65ex\hbox{$\sim$} }}

\begin{document}
\title{Evolution of globular cluster systems in elliptical
galaxies. I: log-normal initial mass function}

\author[E.Vesperini]{E.Vesperini\thanks{E-mail:vesperin@falcon.astro.umass.edu
}
\\ Department 
of Astronomy, University of Massachusetts, Amherst, MA, 01003, USA}
\maketitle
\begin{abstract}
We study the evolution of globular cluster systems (GCS) in elliptical
galaxies and  explore the dependence of their main  
properties on the mass and the size of the host galaxy. 

We have focussed our attention on the evolution of the GCS mass
function (GCMF), on the fraction of surviving clusters and on the
ratio of the final to initial total mass in clusters; the
dependence of these GCS properties on the structure of the host galaxy
as well as their variation  
with the galactocentric distance inside individual host galaxies has
been thoroughly investigated.  We have adopted a log-normal initial
GCMF with mean mass and dispersion ($\lm_i=5.25$ and $\sg_i=0.6$)
similar to those observed in the external regions of elliptical
galaxies where memory of initial conditions is likely to be well preserved.

After a survey over a large number of
different host galaxies we have restricted our attention to a sample
of  galaxies with effective masses, $M_e$, and radii,
$R_e$, equal to those observed for dwarf, normal and giant
ellipticals. 

We show that, in spite of large differences in the fraction of
surviving clusters,
the final mean masses, $\lmf$,  of the GCMF in massive galaxies ($\log M_e
\gtorder 10.5$) are very similar to each other ($\lmf\simeq 5.16$,
$M_V=-7.3$ assuming $M/L_V=2$)
with a small galaxy-to-galaxy dispersion; low-mass compact galaxies
tend to have smaller values 
of $\lmf$ and a larger galaxy-to-galaxy dispersion. These findings are
in agreement with those of recent observational analyses. 

The fraction of surviving clusters, $\nfni$, increases with the mass
of the host galaxy ranging from $\nfni \sim 0.9$ for the most massive
galaxies to $\nfni \sim 0.1$ for dwarf galaxies.
We show that a small difference between the initial and the final mean
mass and dispersion of the GCMF and the lack of a significant radial
dependence of $\lmf$ inside individual galaxies do not necessarily
imply that evolutionary processes have been unimportant in the
evolution of the initial population of clusters. For giant galaxies
most disruption occurs within the effective radius while for low-mass
galaxies a significant disruption of clusters takes place also at
larger galactocentric distances. 

The dependence of the results obtained on the initial mean mass of the
GCMF is also investigated and it is shown that outside the interval
$4.7\ltorder \lmi \ltorder 5.5$ both the spread and the numerical
values of $\lmf$ are not consistent with those observed. 
\end{abstract}
\section{Introduction}
The study of the properties of  globular cluster systems (hereafter
GCS)  in our Galaxy and in external galaxies  can
provide important clues on the formation and the evolution of individual
globular clusters and of their host galaxies. A large number of  observational
investigations have been carried out (see e.g. Ashman \& Zepf 1998 and Harris
2000 for recent reviews)  and  the wealth of
data collected have provided several indications on the
relations between the properties of GCS  and those of the host galaxies as
well as on the dependence of the properties of individual GCS on the position
inside their host galaxies. 

On the theoretical side, a firmly established theory for the formation of 
globular clusters is still lacking and the interpretation of a number of
observational results is matter of debate. Several theoretical studies
addressing the evolution of the properties of GCS have concluded that
evolutionary processes lead to the disruption of a significant
number of globular clusters  and can change the shape and the
parameters of the initial mass function of the system (see e.g.
Fall \& Rees 1977, Fall \& Malkan 1978, Caputo \& Castellani 1984, Chernoff,
Kochanek, Shapiro 1986, Chernoff \& Shapiro 1987, Aguilar, Hut \&
Ostriker  1988,
Vesperini 1994, 1997, 1998, Okazaki \& Tosa 1995, Capuzzo Dolcetta \&
Tesseri 1997,  Gnedin \& Ostriker 1997,
Murali \& Weinberg 1997a, 1997b, Ostriker \& Gnedin 1997, Baumgardt
1998; see also Meylan \& Heggie 1997 for a recent review on the
dynamical evolution of globular clusters).  

On the other hand, many observational 
investigations have  shown that the mass functions of globular cluster
systems (hereafter we will indicate the mass function of a globular
cluster system by GCMF and the luminosity function of a globular
cluster system by GCLF) of
galaxies with structures markedly different from each other, in which
the efficiency of
evolutionary processes should be different, are very similar to each
other  and they
are all well fitted by a log-normal function 
with approximately the same mean value and dispersion (see e.g. Harris
1991); the turnover of the GCLF has been often used as a standard
candle calibrated on the value of 
the turnover of GCS of galaxies in the Local Group  for the
determination of extragalactic distances (see e.g. Jacoby et al. 1992).

As to the 
radial variations of the GCMF properties inside individual galaxies, while in
some  observational analyses 
it has been claimed that inner clusters tend to be 
more luminous than outer clusters (Crampton et al. 1985 for the M31
GCS, van den Bergh 1995a, 1996 for the Galactic GCS, Gnedin 1997 for the GCS
of the Milky Way, of M31 and of M87, see also Ostriker \& Gnedin 1997
for a theoretical analysis of the data presented in Gnedin 1997; 
see also Chernoff \& Djorgovski 1989, Djorgovski \&
Meylan 1994 and Bellazzini
et al. 1996 for evidence of differences in the structure of inner and
outer Galactic clusters), most observational studies fail to
report any significant radial trend in the properties
of the GCMF (see e.g. Forbes et al. 1996, Forbes, Brodie \& Hucra 1996, 
Forbes, Brodie \& Hucra 1997, 
Kavelaars \& Hanes 1997, Harris, Harris \& McLaughlin 1998, Kundu et
al. 1999; see also Gnedin 1997 for a discussion of the possible role of
the statistical methodology used in the determination of the difference
between the GCLF parameters of inner and outer clusters).

The apparent universality of the GCMF in galaxies with different structures and
the lack of a strong radial gradient of the properties of the GCMF inside
individual galaxies have been often interpreted, at odds with the
conclusions of the theoretical studies mentioned above, as an indication that
evolutionary processes do not play a relevant role in determining the current
properties of GCS. 

In some recent
investigations the issue of the universality of the GCMF has been
addressed more in detail (Whitmore 1997, Ferrarese et al. 2000,
 Harris 2000) and the 
absolute magnitude of  the turnover of the GCLF have been determined
using primary standard candles; these studies have shown that, in
fact, there is an intrinsic 
scatter in the properties of GCMF of galaxies of the same type and
that there are some
non-negligible differences between the mean properties of the GCMF of
galaxies of different type.  In particular the results of the most recent of
these studies (Harris 2000) show that 1) giant ellipticals have a
mean turnover magnitude, $M^0_V$, equal to -7.33 with a galaxy-to-galaxy rms
scatter equal to 0.15 (corresponding, for $M/L_V=2$, to $\lm=5.16$ and
a scatter of 0.06), 2) dwarf ellipticals have a mean turnover about
0.4 mag fainter than 
giants, $M_V^0=-6.9$ with a galaxy-to galaxy rms scatter of 0.6 mag
(corresponding to $\lm=4.99$ and a scatter of 0.24),  
3) for disk galaxies, the mean turnover of the GCLF, $M_V^0=-7.46$,
appears to be about 0.15 mag brighter than that of giant ellipticals,  with a
dispersion of about 0.2 mag ($\lm=5.21$ and a scatter of 0.08).

In this paper we will focus our attention on elliptical galaxies and we will
explore the dependence of the evolution of the GCMF, of the  fraction
of clusters surviving after one Hubble time and of the ratio of the
final to the initial total
mass of clusters   on the  structure of the host
galaxy; the efficiency of evolutionary processes in producing 
a significant radial gradient of the properties of the GCMF inside individual
galaxies will be studied. 
The results of our simulations will be compared with the
observational data available for giant, normal and dwarf elliptical galaxies.

We will adopt a log-normal initial GCMF with parameters equal to
those observed  in the external regions of some galaxies where evolutionary
processes are unlikely to have altered the initial
conditions of 
the GCS; we will not consider here younger globular cluster
populations which could have formed during mergers or interactions. In
a companion paper (Vesperini 2000) we will consider the 
evolution of GCS with  a power-law initial GCMF similar to that of young
cluster systems observed in merging galaxies.

The layout of the paper is the following:
in \S 2 we describe the method adopted for this investigation; 
in \S 3 we  explore
the evolution of GCS located  in a large set of  host
galaxies with different effective masses and radii and derive a number
of general results on the dependence of 
the evolution of GCS on the host galaxy; then we 
show the implications of our results for  a sample of host galaxies
with effective masses and radii equal to those estimated
observationally for a number of ellipticals; the results obtained are
compared with the observed trends in the 
properties of GCS in elliptical galaxies. In \S 4 we study the
dependence of our results on the initial mean 
mass of the GCMF. In \S 5 we discuss and summarize all the results.
\section{Method}
In order to calculate
the evolution of the masses of individual globular clusters in a GCS, we will 
adopt the expressions derived by means of large set of N-body simulations by
Vesperini \& Heggie (1997) and already used to study the evolution of the
Galactic GCS in Vesperini (1998). 
The evolutionary
processes we will consider are the mass loss associated to the evolution of
individual stars in a cluster, two-body  relaxation, the presence of the
tidal field of the host galaxy and dynamical friction. The effects due
to the time variation of the tidal field for clusters on elliptical orbits (see
e.g. Weinberg 1994a,b,c, Gnedin, Hernquist \& Ostriker 1999) were not
considered in the simulations of Vesperini 
\& Heggie and are not included here.  
For details on the N-body simulations we refer to Vesperini \& Heggie (1997). 
\begin{figure}
\centerline{\psfig{file=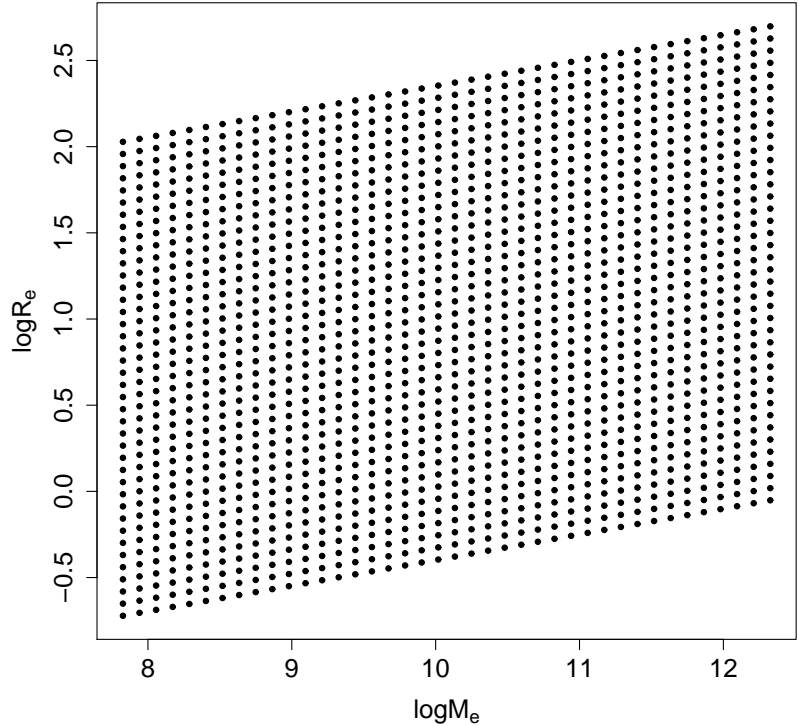,height=11cm,width=11cm,angle=0}}
\caption{Set of values of the effective mass, $M_e$ ($M_{\odot}$), and of the
effective radius, $R_e$ (kpc), of the host galaxies considered in the paper.}
\end{figure}

 Fitting the time evolution
of the mass of individual clusters as obtained by N-body simulations, Vesperini
\& Heggie (1997) derived the following analytical expression for a
cluster with  
initial mass $M_i$ and located at a galactocentric distance $R_g$
\be
{M(t)\over M_i}=1-{\Delta M_{st.ev.}\over M_i}-{0.828\over F_{CW}}t
\ee
where $t$ is time measured in Myr, ${\Delta M_{st.ev.}\over M_i}$ is the mass
loss due to stellar evolution (see eq.10 in Vesperini \& Heggie 1997) and
$F_{CW}$ is a parameter introduced by Chernoff \& Weinberg (1990)
which is proportional to the initial relaxation time of the cluster and is
defined as  
\be
F_{CW}={M_i \over M_{\odot}}{R_g \over \hbox{kpc}}{1\over \ln N}{220
\hbox{km s}^{-1} \over v_c},
\ee
where $N$ is the total initial number of stars in the cluster and $v_c$ is the
circular velocity around the host galaxy. 

For the host galaxy we will assume a simple isothermal model with constant
circular velocity.

The effects of dynamical friction at any time $t$ are included by removing, at
that time, all clusters with time-scales of orbital decay (see
e.g. eq. 7.26 in  Binney \&
Tremaine 1987) smaller than $t$.

Each  host galaxy we have considered is characterized by a
value of the effective radius, $R_e$, and of the effective mass $M_e$.
For each GCS we have explored, we have drawn 20000 random values of
$M_i$ according to the initial GCMF chosen and the distances of clusters from
the center of the host galaxy are such that the number of cluster  per
cubic kpc 
is proportional to $R_g^{-3.5}$ with $R_g$ ranging from $0.16R_e$ to
$5R_e$. The adopted profile for the radial distribution is similar to
that observed for Galactic halo clusters; a similar slope for the
initial radial distribution has been obtained by Murali \& Weinberg
(1997a) from detailed models of the evolution of the GCS of M87. 

We have studied the evolution of GCS in a
large set of different host galaxies:  Fig. 1 shows all the pairs
$(\log M_e, \log R_e)$ considered. 

The evolution of each GCS has been followed for one
Hubble time here taken equal to 15 Gyr.
\section{Results}
\subsection{General results}
In this section we discuss the results obtained adopting a log-normal  initial
GCMF with mean mass $\lmi=5.25$ and dispersion $\sg_i=0.6$ which is
similar, for 
example, to the GCMF of outer (and thus unlikely to be significantly
affected by evolutionary processes) clusters in M87 (see
e.g. McLaughlin, Harris \& Hanes 1994, Gnedin 1997). 
\begin{figure}
\centerline{\psfig{file=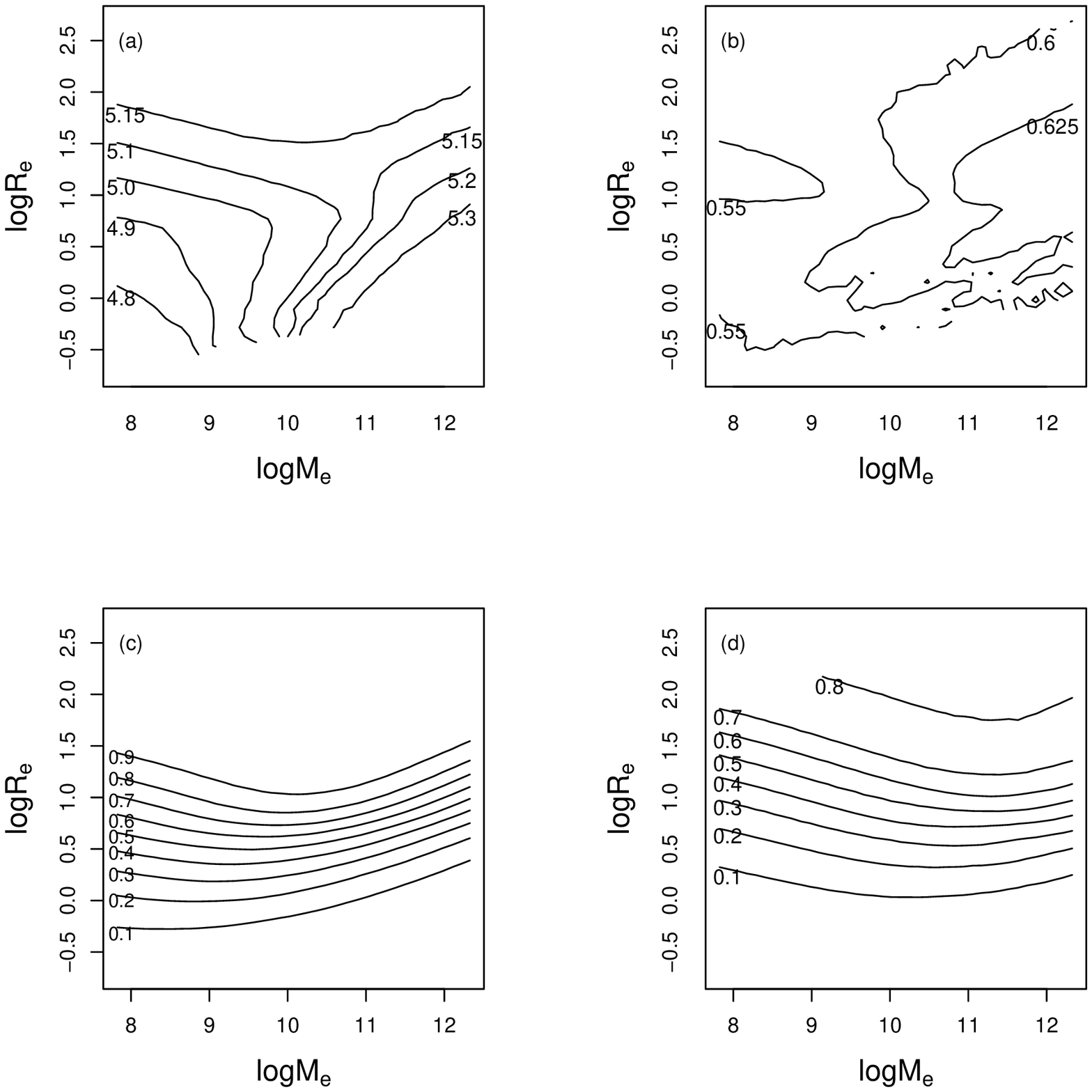,height=11cm,width=11cm,angle=0}}
\caption{Contour plots of (a) $\lmf$, (b) $\sgf$, (c) $\nfni$, (d)
$\mfmi$ in the plane $\log M_e-\log R_e$.}
\end{figure}
Figs 2a and 2b show the contour plots
of the final (at $t=15$ Gyr) mean mass and dispersion of the GCMF,
$\lmf$ and $\sgf$,  in the plane 
$\log M_e-\log R_e$.  In most host galaxies with low values of $M_e$, 
for which dynamical friction is more important than evaporation,
$\lmf<\lmi$ while $\lmf>\lmi$ for high-mass host galaxies where
mass loss and disruption due to two-body relaxation are more important.
 It is interesting to note that for a large number of host galaxies,
$\lmf$  and $\sgf$ do not differ significantly from their initial
values  
\footnote{In comparing initial and final values of
$\lm$, one has to consider that, no matter how efficient dynamical friction and
evaporation are, all GCS have their initial total mass and mean mass 
decreased because of the mass loss associated to stellar evolution;
for the initial 
stellar mass function used in  the simulations by Vesperini \& Heggie
(1997), on 
which this investigation is based, after 15 Gyr the mass lost by each cluster
because of stellar evolution amounts to about 18 per cent  of the
initial total mass; this means that also in those GCS in which no
cluster is disrupted, the 
final total mass and the final mean mass are always smaller than or equal to
approximately 0.82 times the corresponding initial values.}. 
 Figs 2c and 2d show the contour plots of the ratio
of the final to the initial total number of clusters, $N_f/N_i$, and of the 
ratio of the final to the initial total mass of clusters, $\mfmi$. The
four panels of Fig. 2 clearly show that {\it a small 
difference  between the initial and the final parameters of the GCMF  does not 
necessarily imply that the number of clusters disrupted by evolutionary
processes  is negligible}. As already shown in Vesperini
(1998) for the Galactic globular cluster system, the gaussian
shape is always well preserved during evolution and the disruption of a large
number of clusters does not always lead to a strong variation in the
values of the initial mean mass and dispersion. 
\begin{figure}
\centerline{\psfig{file=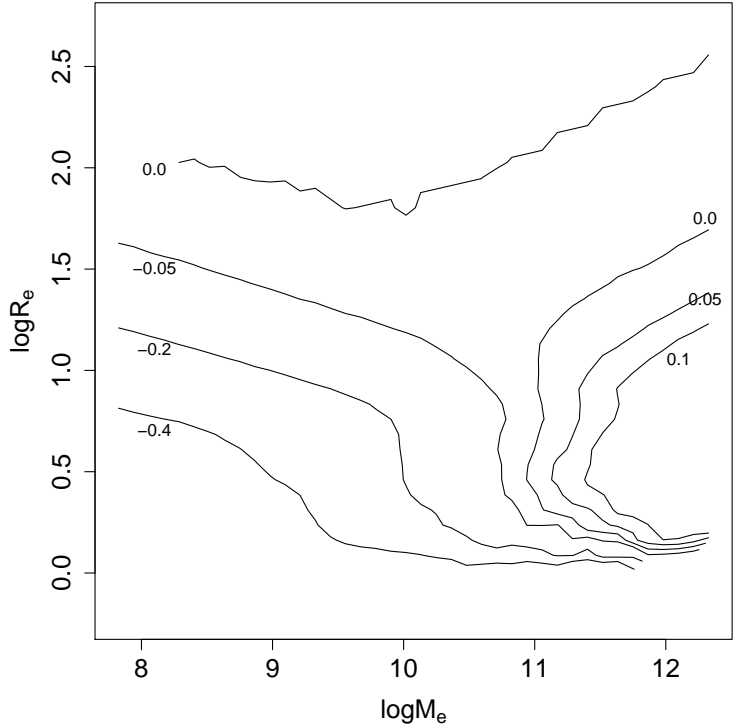,height=11cm,width=11cm,angle=0}}
\caption{Contour plot of the difference between the final mean
mass of inner clusters ($R_g<R_e$) and of outer clusters ($R_g>R_e$), $\dmio$,
in the plane $\log M_e-\log R_e$.}
\end{figure}

The minima of the curves of constant
$N_f/N_i$ and $\mfmi$ approximately correspond  to the transition from a regime
dominated by evaporation to one dominated by dynamical friction: if only
the effects of evaporation were  taken into account, the curves 
of constant $N_f/N_i$ and $\mfmi$ would go down from
the upper right to the lower left region of the $\log M_e-\log R_e$
plane;  as
dynamical friction takes over as the dominant evolutionary process, the curves
of constant $N_f/N_i$  and $\mfmi$ turn up leading to values of these
quantities smaller than those one would have if evaporation were the only
process included.

Fig. 3 shows the contour plot of the difference, $\dmio$, 
between the mean mass of inner clusters (defined as those with galactocentric
distance smaller than $R_e$) and outer clusters (those located at 
galactocentric distances larger than $R_e$). 

Since the efficiency of evolutionary
processes depends on the galactocentric distance, the formation of a radial
gradient in the parameters of the GCMF has to be expected; nevertheless, as
shown in Fig.3, in several cases this gradient is very
weak. Moreover, since the observed positions  of globular
clusters in external galaxies are not real galactocentric distances but
projected distances,  any radial gradient will appear even
weaker and its detection is likely to be difficult (see
e.g. Harris 2000).  
\begin{figure}
\centerline{\psfig{file=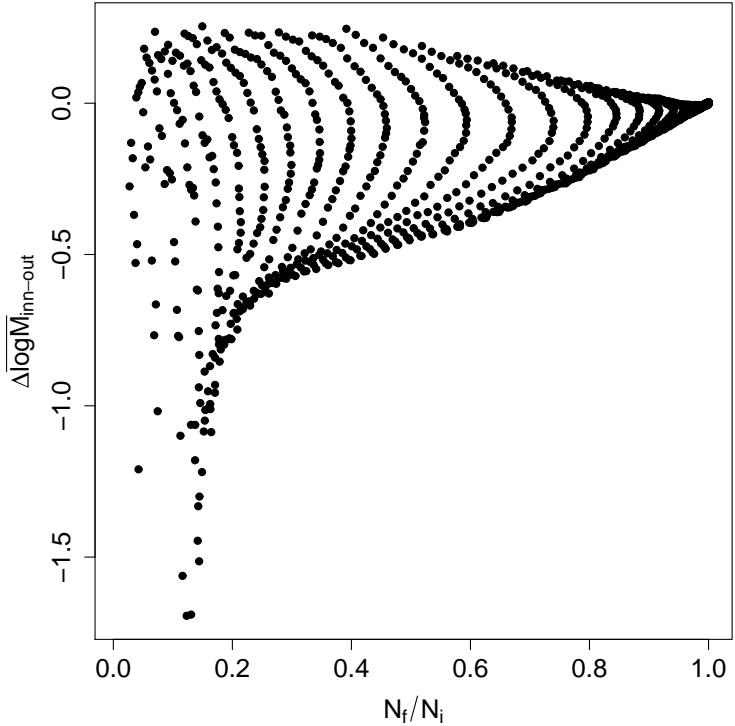,height=11cm,width=11cm,angle=0}}
\caption{Difference between the final mean
mass of inner clusters ($R_g<R_e$) and of outer clusters ($R_g>R_e$), $\dmio$,
versus the fraction of surviving clusters after 15 Gyr, $\nfni$.}
\end{figure}

Inner clusters are those more affected by
evolutionary processes: depending on whether the dominant evolutionary process
is dynamical friction or evaporation and on the balance between disruption of
clusters by these evolutionary processes and the evolution of the masses of the
clusters which survive (see \S 3.1 in Vesperini 1998 for a detailed
discussion  of the   effects of  disruption and evolution of the masses of
surviving clusters) $\dmio$ can be positive or negative and, as shown
in Fig. 3, in a number of cases  it is very close to zero.  Another
point to note is that, in principle, the 
radial gradient of $\lm$, if investigated in finer radial bins, does
not need to be monotonic: it is conceivable that as
$R_g$ increases, the dominant evolutionary process and the balance between 
disruption and evolution of the masses of the surviving clusters which are
responsible for the evolution of $\lm$, can change  in a way leading
to a non-monotonic behavior of $\lm$. 

Fig. 4 shows $\dmio$ versus the
fraction of surviving clusters, $N_f/N_i$:  it is important to remark that,
while all GCS characterized by a large  value of
$\dmio$ have had their initial population of clusters significantly
depleted  by evolutionary processes,  {\it a weak or negligible radial gradient
of the  mean mass of the GCMF does not imply that  evolutionary processes have
not played an important role  in modeling the properties of individual
clusters and in disrupting a significant fraction of the initial number of 
clusters}; as Fig. 4 shows, there are a large number of systems in which much
less than 50 per cent of the initial number of clusters have survived
but still $\dmio$ is equal to or close to zero.
\subsection{Implications}
We now turn our attention to the implications of the results discussed
above for a sample of host galaxies with structural properties known
from observations. 
We will refer to the sample of elliptical galaxies for which the values of
$M_e$ and $R_e$ are provided in Burstein et al. (1997). Here we will adopt 
values of $M_e$ and $R_e$ corresponding to $H_0=75~\hbox{km
s}^{-1}\hbox{Mpc}^{-1}$.  Figs 5a-d show the contour plots of
$\lmf$, $\sgf$, 
$N_f/N_i$ and $\mfmi$ in the plane $\log M_e-\log R_e$  with the
observational values of
$\log R_e$ and $\log M_e$  superimposed; these figures
illustrate the expected final properties of the GCS of the galaxies
considered. 
\begin{figure}
\centerline{\psfig{file=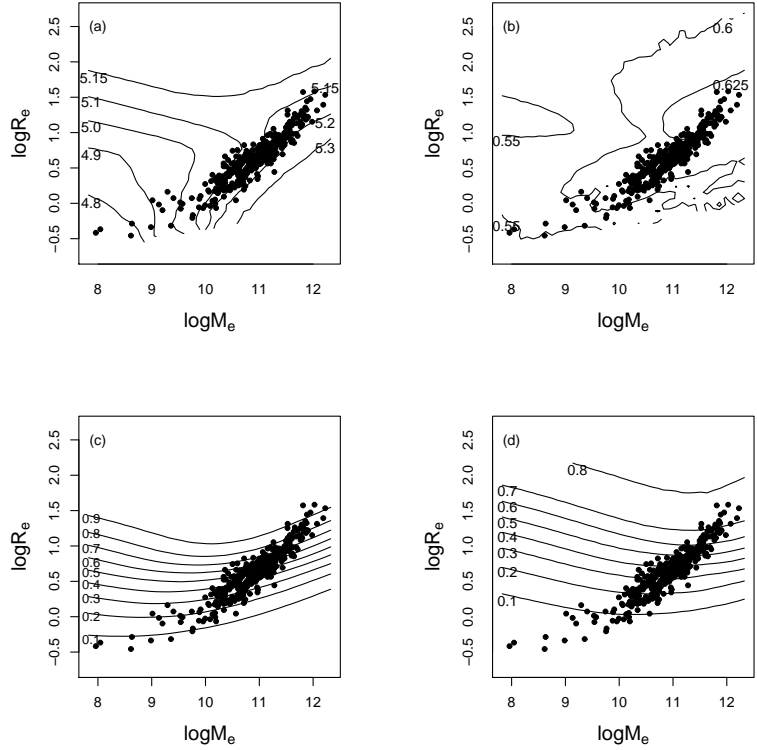,height=11cm,width=11cm,angle=0}}
\caption{Contour plots of (a) $\lmf$, (b) $\sgf$, (c) $\nfni$, (d)
$\mfmi$ in the plane $\log M_e-\log R_e$ (already shown in Figure 2) with the
observational values of $\log M_e$ and $\log R_e$ for elliptical galaxies 
(data from Burstein et al. 1997) superimposed as filled dots.}
\end{figure}

The fraction of the initial population of clusters which survive
after one Hubble time and the ratio  of the final to initial total
mass in clusters  
for the sample of galaxies considered span the entire
range of possible values; in contrast with such a broad range of
values of $N_f/N_i$ and $\mfmi$,  Fig. 5a shows that most galaxies (in
particular those with $\log M_e \gtorder 10$) occupy a region of the
$\log M_e-\log R_e$ plane corresponding to a very narrow range of
$\lmf$: {\it a 
small range of values of $\lmf$ in galaxies with different structures
implies neither that the fraction of the initial number of clusters
which have been disrupted in these galaxies is similar nor 
that evolutionary processes did not alter the initial conditions of individual
clusters}.  This point is further illustrated in Figs 6a-d where we have
plotted  $\lmf$, $\sgf$, $N_f/N_i$ and
$\mfmi$ versus the mass of the host galaxy. 
Fig. 6a
shows that $\lmf$ is approximately constant for 
$\log M_e\gtorder 10-10.5$ while it decreases for smaller galaxies;
this result is  perfectly consistent with the findings
of the observational analysis by  Harris (2000) discussed above in the
Introduction which shows that the mean luminosity of globular clusters is
approximately constant in giant ellipticals and that the mean luminosity
of clusters in dwarf galaxies is fainter than that of giant
ellipticals.

  Figs 6c and 6d show that 
$N_f/N_i$  and $\mfmi$ increase with the mass of the host galaxy. The
trend of $\nfni$ and $\mfmi$ to increase with the mass of the host
galaxy results from the relation between the observational values of $M_e$
and $R_e$ (for $\log M_e \gtorder 10$, $R_e \sim M_e^{0.66}$; see
Burstein et al. 1997 for further details on the fundamental plane
relations for the sample of elliptical galaxies considered here) which
is such that both the efficiency of relaxation and that of dynamical friction 
decrease as $M_e$ increases (for example $R_e \sim M_e^{0.66}$
implies $F_{CW} \sim M_e^{0.49}$). This is clearly illustrated in Fig
5c: the slope of the relation between  $\log M_e$ and $\log R_e$ 
is steeper than the slope 
(referring for example to the high-$M_e$ side of the plot) of the
curves of constant $\nfni$ and, as $M_e$ increases, observational
points in the $\log M_e- \log R_e$ plane cross curves corresponding to
larger and larger values of $\nfni$.
\begin{figure}
\centerline{\psfig{file=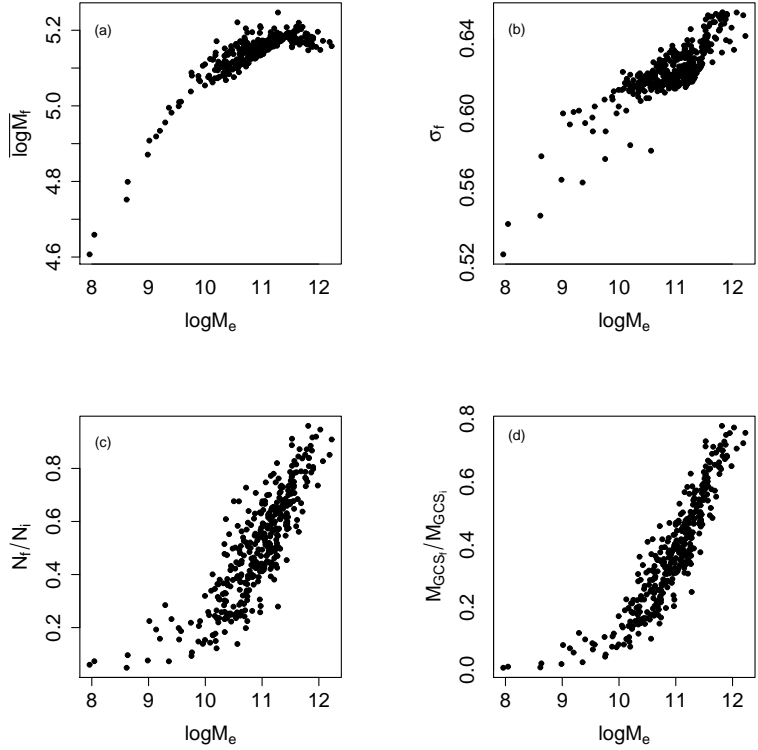,height=11cm,width=11cm,angle=0}}
\caption{(a) $\lmf$, (b) $\sgf$, (c) $\nfni$ and (d) $\mfmi$  from
the simulations discussed in \S 3.2 versus the logarithm of the effective 
mass of  the host galaxy for a set of globular cluster systems located in host
galaxies with values of $R_e$ and $M_e$ equal to the observational values 
plotted in Fig. 5.}
\end{figure}

Fig. 6c is  also important for its implications for the dependence
of the number of clusters on the mass of the host galaxy.  Several
investigations have been carried out to study the  abundance of globular
clusters as a function of the properties of the host
galaxies: the specific frequency,
$S_N$, defined as the number of clusters per unit luminosity (Harris \& van
den Bergh 1981), has been determined for a number of galaxies and, more
recently, Zepf \& Ashman (1993) have introduced a mass-normalized specific
frequency defined as the ratio of the total number of clusters to the
total mass of the host galaxy. In general, observational analyses (see
e.g. Ashman \& Zepf 1998, Elmegreen 2000 and references therein) have
shown that ellipticals 
tend to have specific frequencies larger than spirals and, within the sample of
ellipticals, specific frequency tends to increase with the
luminosity of galaxies (see e.g. Djorgovski
\& Santiago 1992, Santiago \& Djorgovski 1993, Zepf, Geisler \& Ashman
1994, Kissler-Patig 1997). A
similar trend is observed also for the 
mass normalized frequency (see Ashman \& Zepf 1998). For nucleated
dwarf galaxies 
this trend is reversed and the specific frequency tends to increase with
decreasing luminosity of the host galaxy (van den Bergh 1995b, Durrell
et al. 1996, Miller et al. 1998).  
The origin of the observed trend between specific frequency and luminosity (or
mass) of the host galaxy is still not clear 
(see e.g. Elmegreen 2000, Harris 2000 for recent reviews). In a recent
work, McLaughlin (1999; see also Blakeslee 1999)
has shown that the observed trend is consistent with a constant efficiency of
cluster formation  and a varying efficiency of star formation per unit gas mass
available; the trend for nucleated dwarf galaxies could be ascribed,
according to McLaughlin's analysis, to a smaller fraction of gas
retained (and  thus of stars formed) in these galaxies after the
formation of globular clusters. 

Our analysis can provide
information only on the ratio of the final to the initial number of clusters;
only with the additional knowledge of the dependence of $N_i$ on the mass of
the host galaxy it would be possible to determine the dependence of the
current number of clusters on the host galaxy mass. Fig. 6c shows
that the fraction of surviving clusters depends on the mass of the host galaxy;
though with a large scatter, a power-law scaling $\nfni \sim M_e^{0.35}$ fits
well our theoretical results. This implies that for $N_i \sim
M^{\alpha}$ and $M/L \sim L^{\beta}$, $N_f \sim M^{\alpha+0.35}$ or
$N_f\sim L^{(1+\beta)(\alpha+0.35)}$; for $N_i \sim L^{\lambda}$,
$N_f \sim M^{\left({\lambda\over 1+\beta}+0.35\right)}$ or $N_f \sim
L^{0.35(1+\beta)+\lambda}$. 

 We do not aim to a detailed fit of our
 results with the observational correlations but it is interesting to
note that, for example, either assuming 
$N_i \propto M$ or
$N_i \propto L$, our results imply that evolutionary processes will lead to a
dependence of the current specific frequency on the mass (and on the luminosity
as well) of the host galaxy: in particular, for $N_i \propto L$ we obtain
$N_f \sim M^{1.16}$ or, adopting $M/L\sim L^{0.24}$ (see e.g. Faber et
al. 1987), $N_f \sim L^{1.43}$; for $N_i  
\propto M$ we obtain $N_f \sim M^{1.35}$ or  $N_f \sim L^{1.67}$. We
thus conclude 
that, starting with a mass- (or luminosity-) independent specific frequency,
evolutionary processes can lead to a trend mass(or luminosity)-specific
frequency consistent with that observed. The possible role of
evolutionary processes in producing a luminosity-specific frequency
correlation has been  pointed out also by Murali \& Weinberg (1997a) and this
result is in agreement with their conclusion. 

It is important to remark that our analysis shows that evolutionary processes
can lead to a trend mass(or luminosity)-specific frequency  without
producing, at the same time, any correlation (which would be in contrast
with observations) between the
mean mass of the GCMF and the mass of the host galaxy 
for galaxies with $\log M_e \gtorder 10.5$. 
\begin{figure}
\centerline{\psfig{file=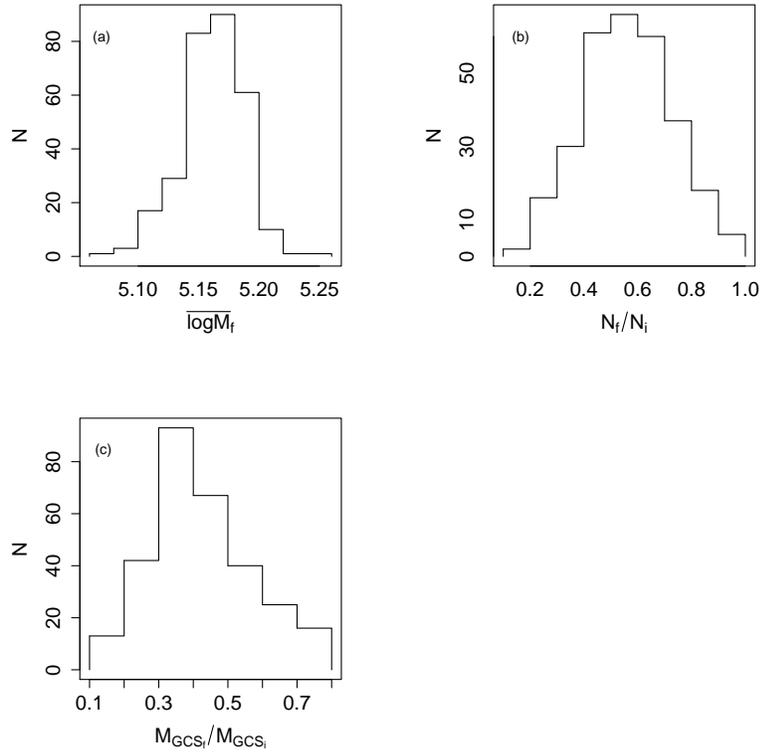,height=11cm,width=11cm,angle=0}}
\caption{Distribution of (a) $\lmf$, (b) $\nfni$ and (c)
$\mfmi$ from the simulations for globular clusters systems located in
host galaxies with values  of $R_e$ and $M_e$ equal to the
observational values plotted in Fig. 5 and having $\log M_e >10.5$.}
\end{figure}

Figs 7a-c show the histograms of $\lmf$,
$N_f/N_i$ and $\mfmi$ for galaxies with $\log M_e >10.5$: the average value
of  $\lmf$ for these galaxies  is equal to 5.16  with a galaxy-to-galaxy rms
dispersion of about 0.03; these values are in good agreement with those found 
by Harris (2000) for giant galaxies. It is interesting to emphasize
again the contrast between the relatively narrow distribution of
$\lmf$ and the much broader distributions of $N_f/N_i$ and $\mfmi$.  

In order to illustrate the
evolution of the distribution of $\lm$ for galaxies with $\log M_e >10.5$  we
have plotted in Fig. 8 the time evolution of the values of $\lm$
corresponding to the 10th, 50th and 90th percentiles of the $\lm$ distribution.
The initial rapid decrease of $\lm$ is caused by the mass loss due to stellar
evolution which does not depend on the host galaxy; the subsequent evolution,
which is essentially determined by two-body relaxation and dynamical friction,
depends on the structure of the host galaxy and
  broadens the distribution of $\lm$ with a slight increase of the
median $\lm$.  
\begin{figure}
\centerline{\psfig{file=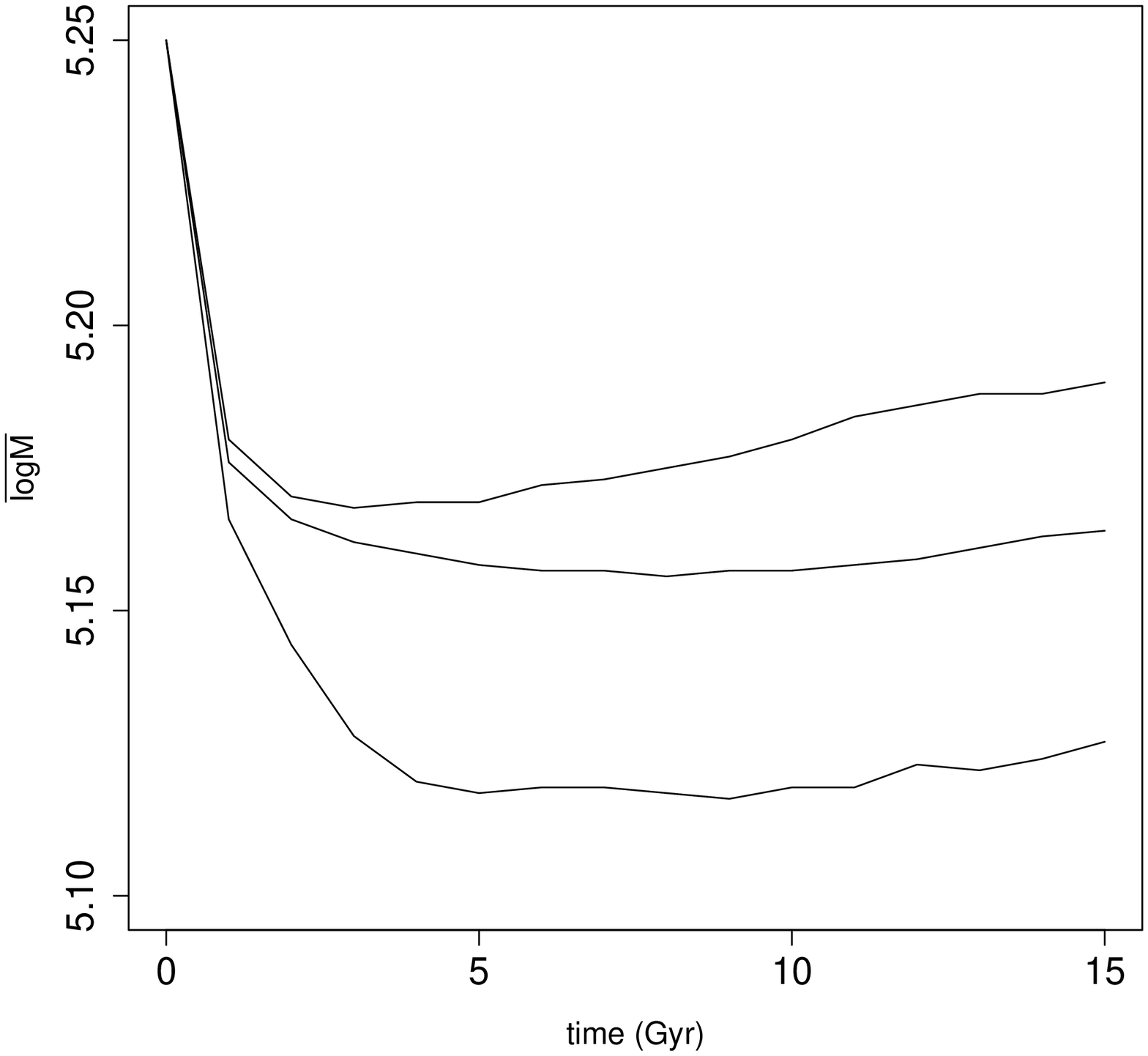,height=11cm,width=11cm,angle=0}}
\caption{Time evolution of the  mean mass
corresponding to the 10th(lower  curve), 50th (central curve) and
90th (upper  curve) percentiles of the distribution of mean masses of
globular clusters systems located in the subsample of host galaxies
with values 
of $R_e$ and $M_e$ equal to the observational values  plotted in Fig. 5 and 
 $\log M_e >10.5$ (see Fig. 7a for the complete distribution of $\lm$
at $t=15$ Gyr for the same sample of host galaxies). }
\end{figure}

For galaxies with $\log M_e <10.5$, plots like those shown in Figs 7 and 8
are less significant because of  the smaller number of objects for which data
are available and of the not uniform distribution of these objects
in the
$\log M_e-\log R_e$ plane. The important point to remark is that,
in general, as is evident from Fig. 6a, there is a clear trend for low-mass
galaxies to have a  
$\lmf$ smaller than that of high-mass galaxies and a larger galaxy-to-galaxy
dispersion; these differences are perfectly consistent with the findings
 of the observational analysis by  Harris
(2000).

In order to further explore the dependence of the properties of the GCMF on
the host galaxy and to better illustrate the differences between
giant, normal and dwarf galaxies 
we have considered the subsample of all the initial
conditions investigated in \S 3.1 corresponding to the region where
most observational data are
located (see Fig. 9). 
\begin{figure}
\centerline{\psfig{file=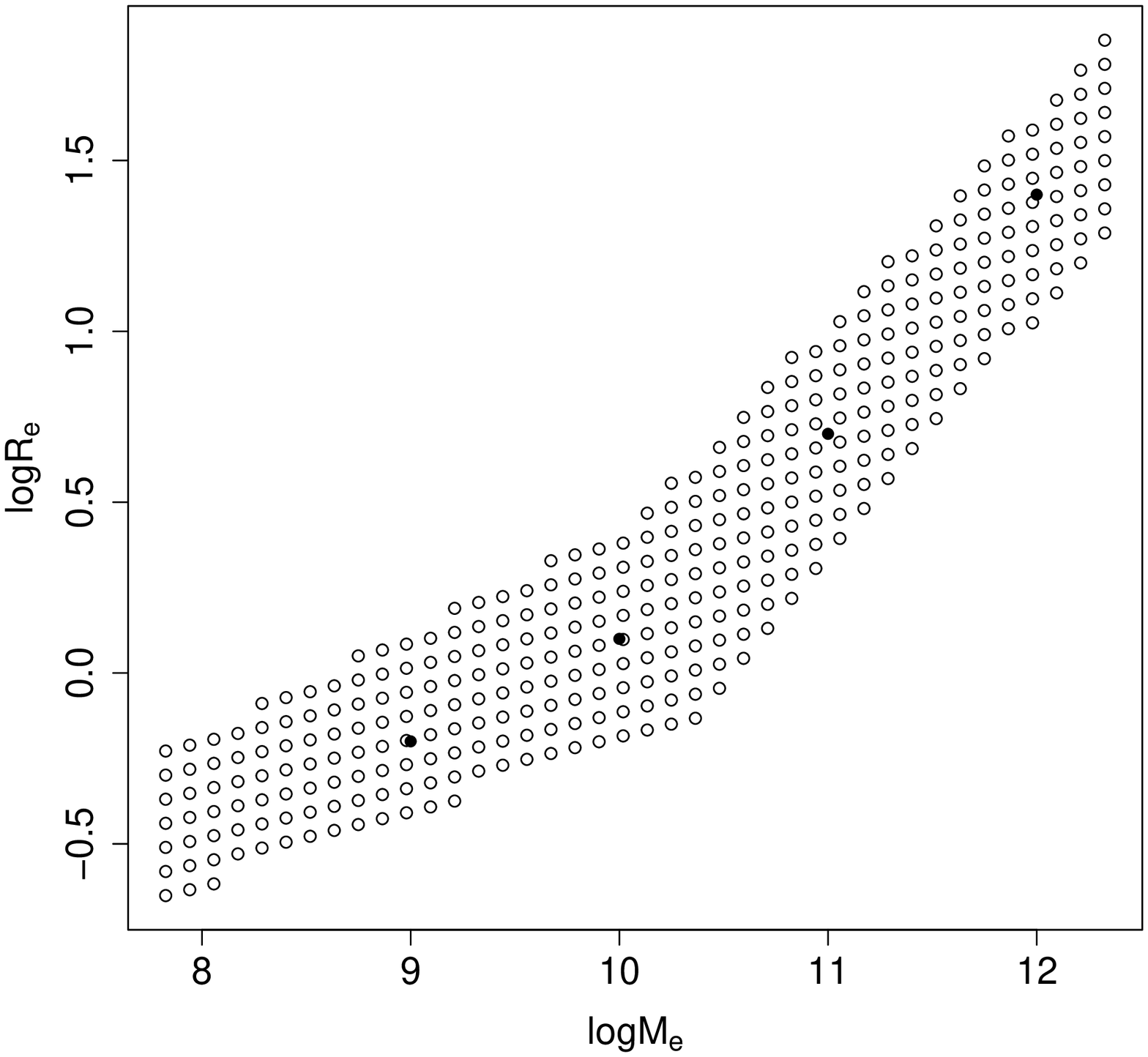,height=11cm,width=11cm,angle=0}}
\caption{ Subsample of values of $R_e$ and $M_e$ considered for the
investigation described in the final part of \S 3.2. 
Filled dots indicate the values of $\log M_e$
and $\log R_e$ for the host galaxies considered in \S 3.3 for a more
detailed study of  the time evolution and of the dependence on the
galactocentric distance of the main GCS properties. }
\end{figure}
For  four
representative cases, shown in Fig. 9 by filled dots, a further detailed
analysis of the time evolution of the properties of the GCMF and of their
dependence on the galactocentric distance has been carried out and the results
are discussed below in \S 3.3.
We have divided the sample of host galaxies  in
three subsamples according to their masses: high-mass galaxies with  
$\log M_e>10.5$, intermediate galaxies with
$9.5<\log M_e<10.5$ and  low-mass galaxies with $\log M_e<9.5$.

The panels  in the left column of Fig. 10 show the distribution of $\lm$ at
$t=2,~5,~10,~15$ Gyr for the three subsamples considered. These plots further
illustrate the expected differences in the mean mass of the GCMF of different
classes of galaxies and the intrinsic galaxy-to-galaxy dispersion in the
distribution of 
$\lm$ in galaxies of the same type; the result discussed
above is confirmed: low-mass galaxies tend to have lower values of
$\lm$ and a larger 
galaxy-to-galaxy dispersion.
\begin{figure}
\centerline{\psfig{file=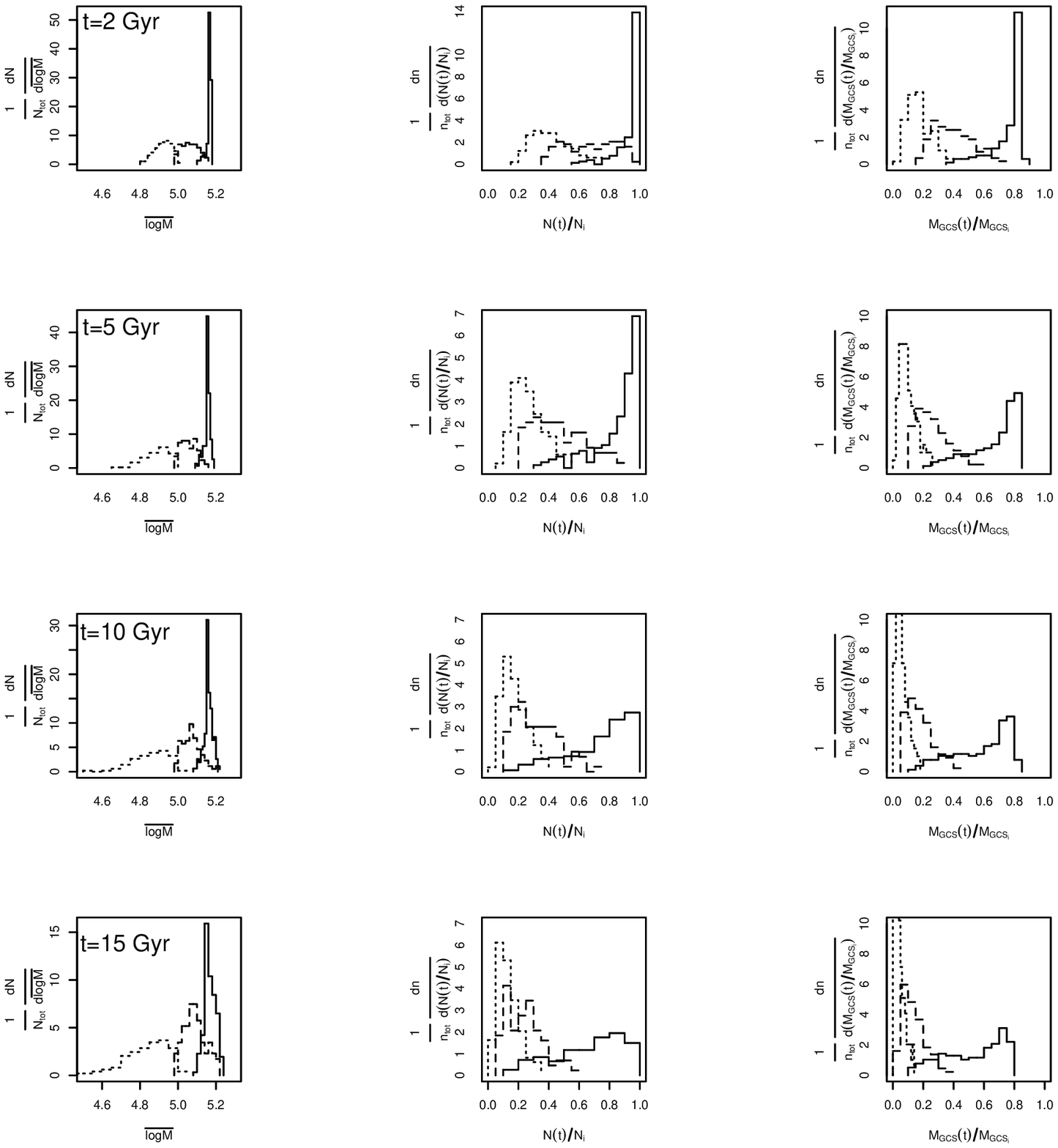,height=20cm,width=20cm,angle=0}}
\caption{ Evolution of the distribution of $\lm$ (left panels),
 of the ratio of the total number of clusters survived at time $t$ to
the total initial number of clusters, $N(t)/N_i$ (central panels), and
of the ratio of the total mass of clusters survived at
time  
$t$ to the total initial mass of all clusters, $M_{GCS}(t)/M_{GCS,i}$, (right
panels) for the set of host galaxies with values of
$R_e$ and $M_e$ plotted in Fig. 9. Each row corresponds to the time
indicated on the first panel on the left. In each panel, the solid line shows
the distribution for globular cluster systems in host galaxies with $\log
M_e>10.5$, the dashed line that for  host galaxies
with $9.5<\log M_e<10.5$ and the dotted line that for host galaxies with $\log
M_e<9.5$. }
\end{figure}

The panels in the central and in the right columns of Fig. 10 show the
distributions of the fraction of surviving clusters and of the ratio
of the total mass in clusters at $t=2,~5,~10,~15$ Gyr to the total
initial mass in clusters for the same samples of host galaxies discussed
above. 
\subsection{Detailed analysis of the evolution of GCS in four fiducial host
galaxies}
In this subsection we focus our attention on 
four fiducial galaxies with values of $\log M_e$ and
$\log R_e$ shown by filled dots in Fig. 9 and we study more in detail
the evolution and the final properties of their GCS.
\begin{figure}
\centerline{\psfig{file=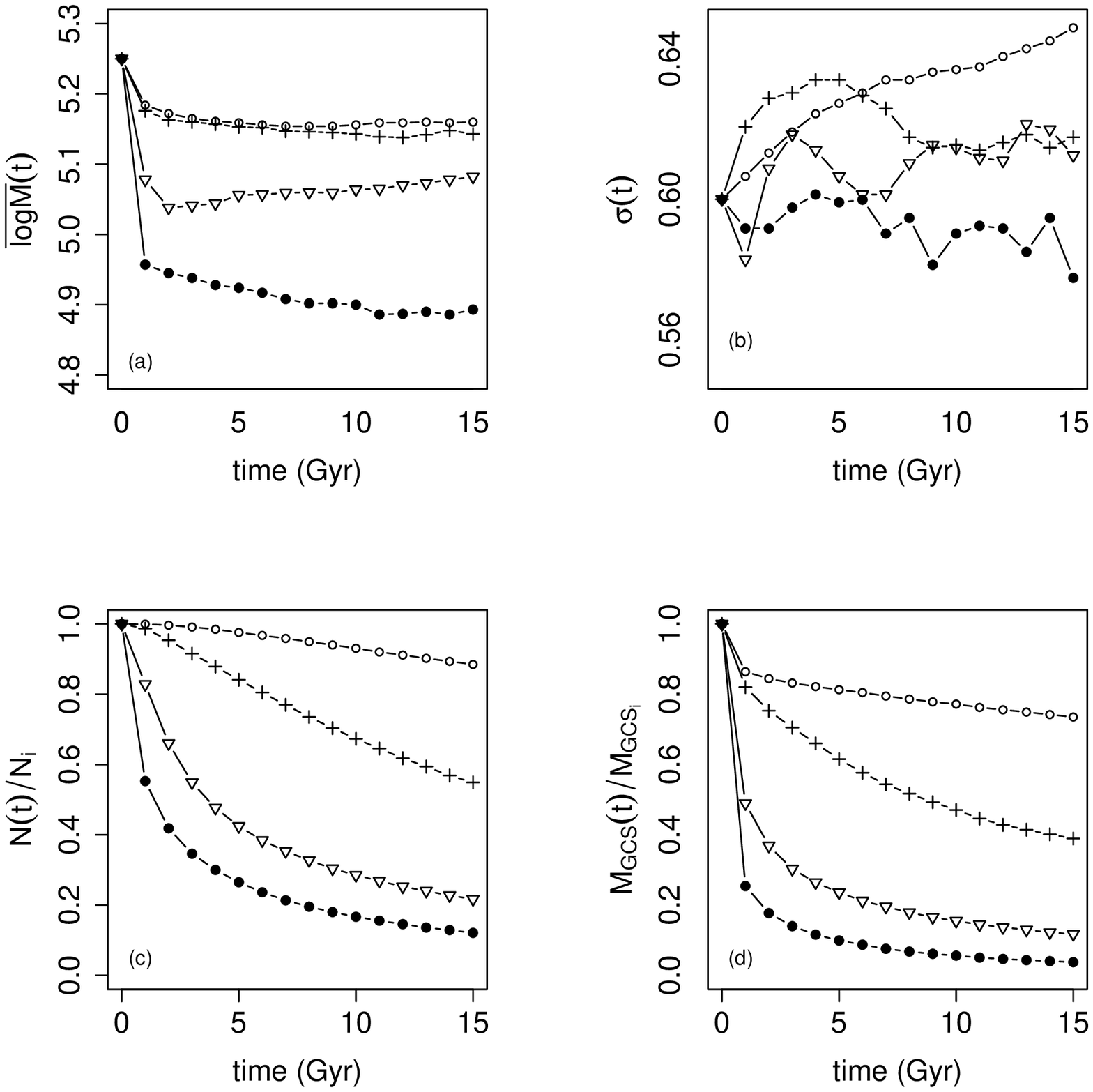,height=11cm,width=11cm,angle=0}}
\caption{Time evolution of the mean mass of the GCMF $\lm$ (a),  of
the dispersion of the GCMF $\sigma$ (b), of  $N(t)/N_i$ (c) and of
$M_{GCS}(t)/M_{GCS,i}$ (d) for globular cluster systems in host galaxies with
effective masses and radii plotted in Fig. 9 as filled dots (open dots
are for $(\log M_e,\log R_e)=(12,1.4)$, crosses for $(\log M_e,\log
R_e)=(11,0.7)$, triangles for $(\log M_e,\log R_e)=(10,0.1)$ and filled
dots for $(\log M_e,\log R_e)=(9,-0.2)$).}
\end{figure}

In Fig. 11 we show the
time evolution of $\lm$ and $\sg$, of the fraction of
surviving clusters and of the ratio of the total mass of surviving clusters to
the total initial mass of clusters. While $\lm$ evolves significantly only 
in the first 2-3 Gyr, individual clusters are continuously disrupted
and those which survive keep losing mass  (see Figs 11c
and 11d): after the initial evolution, the gross properties of the
GCMF adopted  do not evolve significantly despite the continuous
disruption of clusters and the evolution of the masses of the
surviving clusters. 

In Figs 12a-c we plot $\lmf$, $N_f/N_i$ and $\mfmi$  versus the 
galactocentric distance. For the two most massive  galaxies considered,
 evolutionary processes are efficient only within a galactocentric distance
approximately equal to $R_e$ while most clusters survive at larger radii
where most of the  mass loss is that associated to the effects of stellar
evolution. 
\begin{figure}
\centerline{\psfig{file=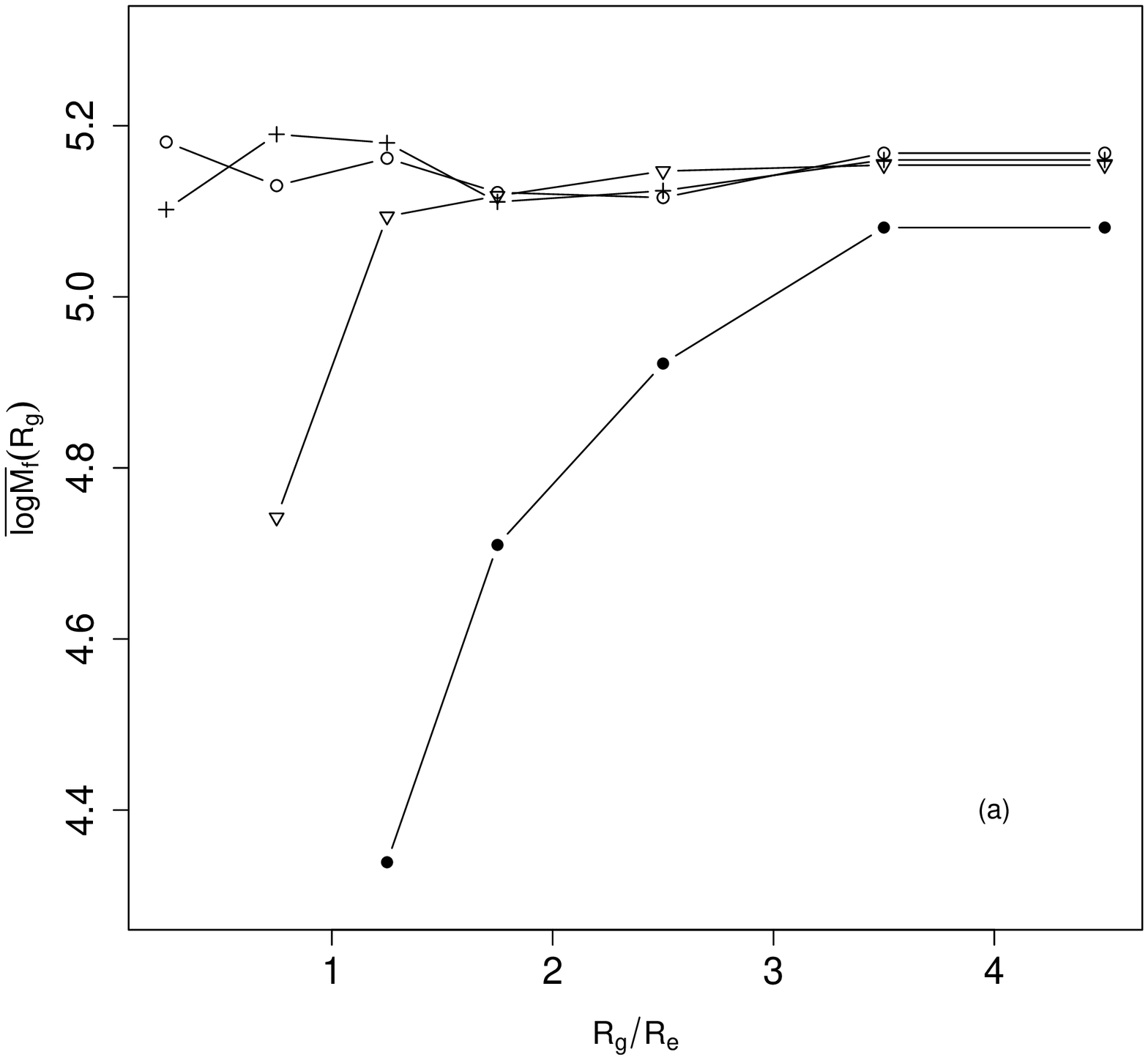,height=7cm,width=7cm,angle=0}}
\centerline{\psfig{file=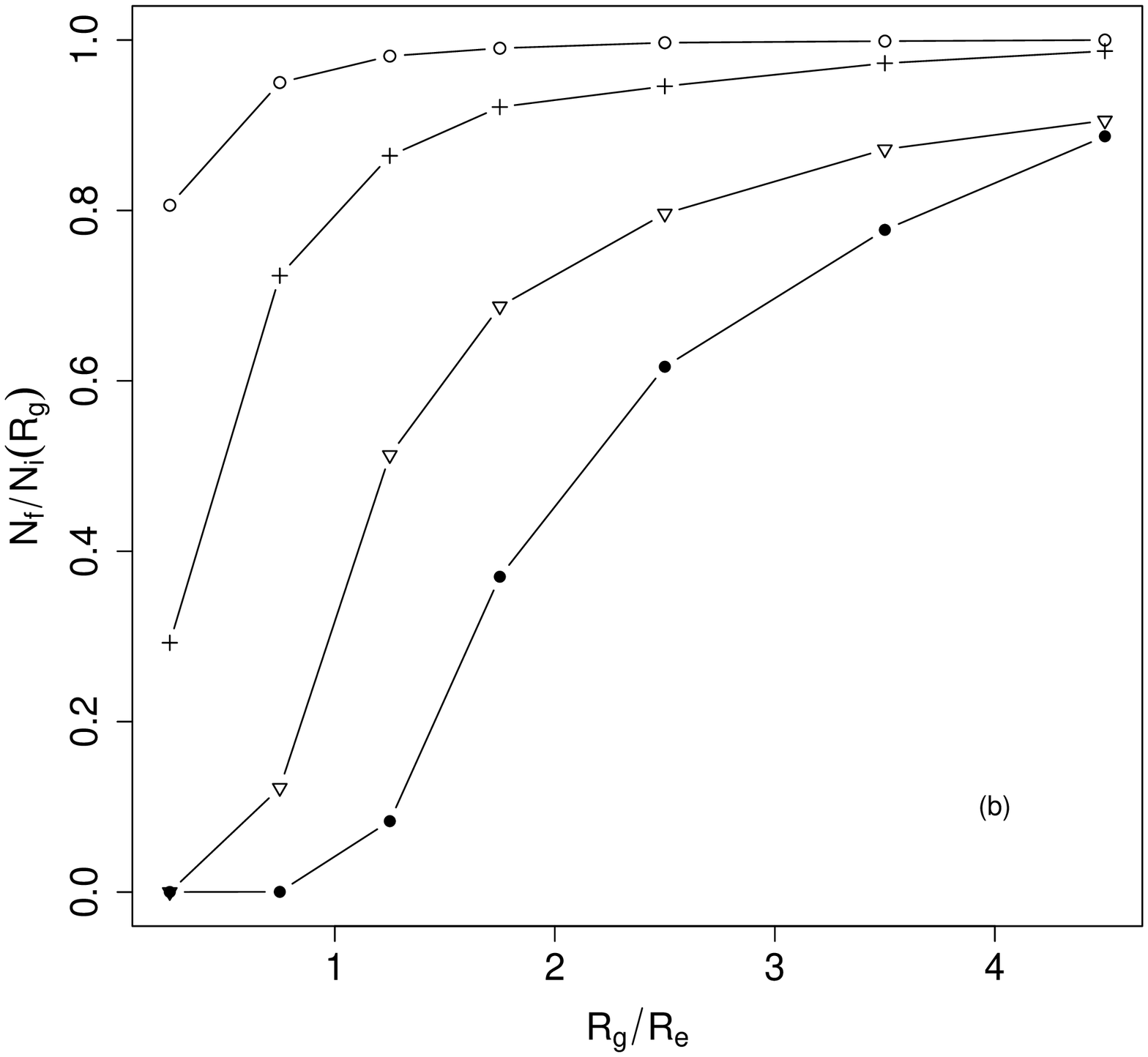,height=7cm,width=7cm,angle=0}}
\centerline{\psfig{file=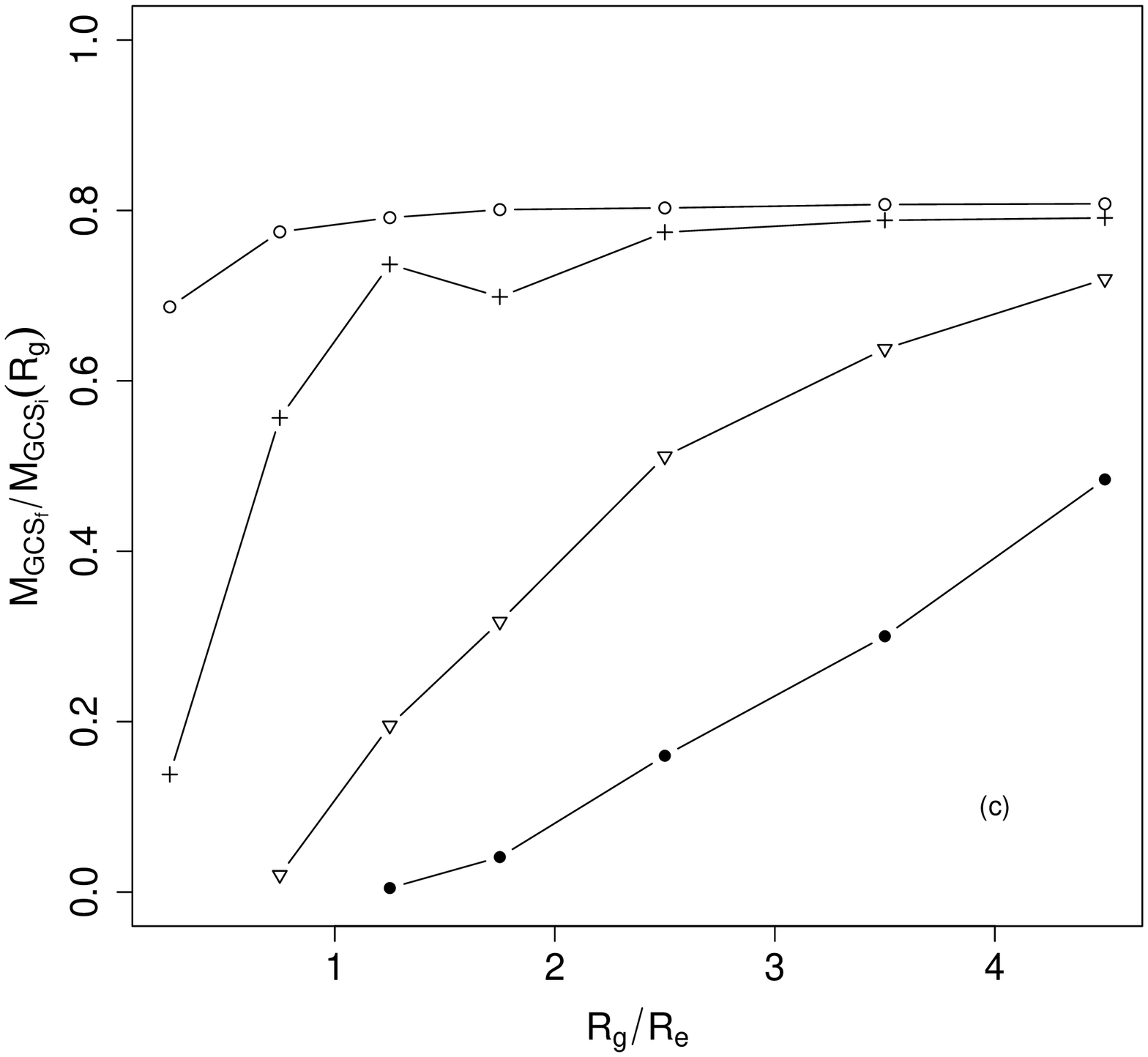,height=7cm,width=7cm,angle=0}}
\caption{(a) $\lmf$, (b)
$\nfni$  and (c) $\mfmi$ versus the  galactocentric distance (normalized by the
effective radius of the host galaxy)  for globular cluster systems in host
galaxies with effective masses and radii plotted in Fig. 9 as filled dots
(symbols for different host galaxies as in Fig. 11).}
\end{figure}
For less massive and more compact galaxies, a strong
disruption and mass loss occur well beyond $R_e$ and even at $R\simeq 2-3 R_e$
about 50 per cent of the clusters are disrupted. In spite of the
significant disruption, only in
the least massive and most compact host galaxy considered, a strong radial
gradient of $\lmf$ extending beyond $R_e$  is produced by the effects of
evolutionary processes. 

 The plot of $N_f/N_i$ vs $R_g$ is of interest 
for the interpretation of the data on the radial dependence of the specific
frequency. Observational data (see e.g. McLaughlin 1999 and references therein)
show that GCS have, in general, a spatial distribution less centrally
concentrated than the stellar halos of their host galaxies with cores larger
than those of the stellar halo while in the outer regions the spatial
distribution of clusters usually matches that of the stellar halo; this implies
that the local specific frequency increases with the galactocentric
distance in the 
central regions and tends to a constant values in the outer parts of the host
galaxy.

Our results show that the increase of the specific frequency with $R_g$ for
$R_g\ltorder 1-2R_e$
can be, at least in part, the result of the effects of
evolutionary processes (see also Capuzzo Dolcetta \& Tesseri 1997,
Murali \& Weinberg 1997a), 
while for $R_g\gtorder 1-2R_e$ evolutionary processes have been 
unimportant in giant galaxies and any observed radial trend has to be
ascribed to the processes of cluster formation. 
Of course, it is not possible to exclude that the decrease
of the specific frequency in the central regions was not in part imprinted by
the formation processes and subsequently further strengthened by evolutionary
processes. 

\section{Dependence of the results on the mean mass of the initial GCMF}
We conclude our investigation 
by exploring the dependence of the main final properties of  GCS on
the initial 
value  of $\lm$ (as to the initial dispersion we will keep the value adopted
until now, $\sigma=0.6$). 

We will restrict our attention to galaxies with $\log
M_e>10.5$. We have adopted the following values of $\log M_e$ and $\log
R_e$: $(\log M_e,\log R_e)=$ (12,1.4), (11.75,1.18), (11.5,1),
(11.25,0.8), (11.0,0.64), (10.75,0.52), (10.5,0.4); 
these values span the entire strip of the $\log M_e-\log R_e$ plane covered by
real galaxies with $\log M_e>10.5$.

  The main  evolutionary process responsible for the disruption of
clusters is evaporation due to internal relaxation for GCS with low values of
$\lmi$, or, as $\lmi$ increases, disruption of high-mass clusters by dynamical
friction. As discussed in Vesperini (1998),  $\lmf$ is
larger or smaller than $\lmi$ depending on the balance between
disruption of low-mass clusters by evaporation, disruption of
high-mass clusters by dynamical friction and  evolution of the masses
of the clusters which survive:  for low values of $\lmi$ most of the
disrupted clusters are low-mass 
clusters and, in general, $\lmf>\lmi$ while, as $\lmi$ increases,
disruption by dynamical 
friction becomes the most important process and most of the disrupted clusters
are those in the high-mass tail of the initial GCMF which leads to
$\lmf<\lmi$.  
 
Fig. 13 shows $\lmf$, $N_f/N_i$ and $\mfmi$ as a function of
$\lmi$ for the seven fiducial host galaxies considered. The range of values
spanned by
$\lmf$ is larger for low and high values of $\lmi$ while for intermediate
values of $\lmi$ ($4.7\ltorder \lmi \ltorder 5.5$), in spite of the
significant spread of values 
of $\nfni$ and of $\mfmi$, the range of values of $\lmf$ is very small.
\begin{figure}
\centerline{\psfig{file=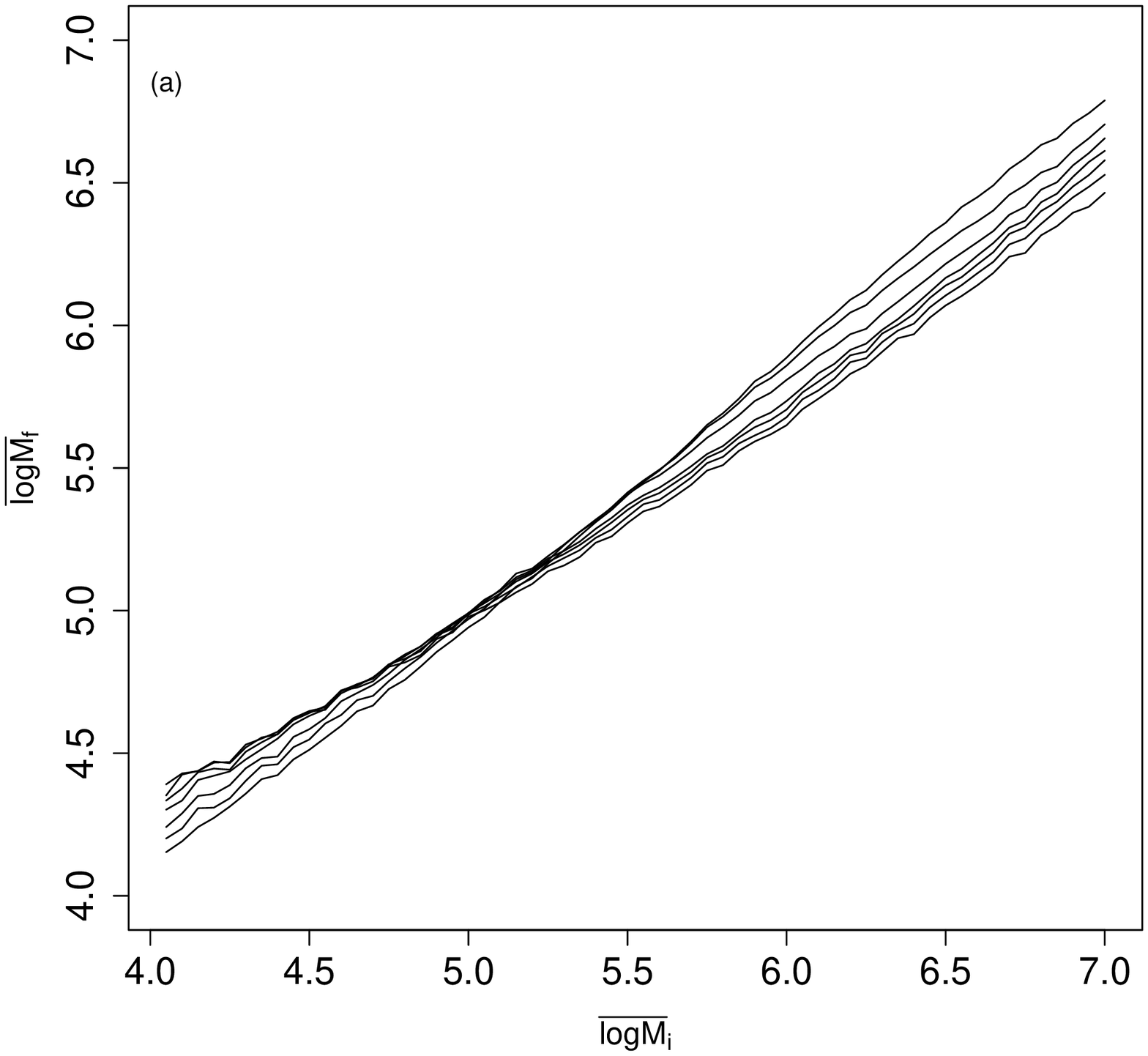,height=7cm,width=7cm,angle=0}}
\centerline{\psfig{file=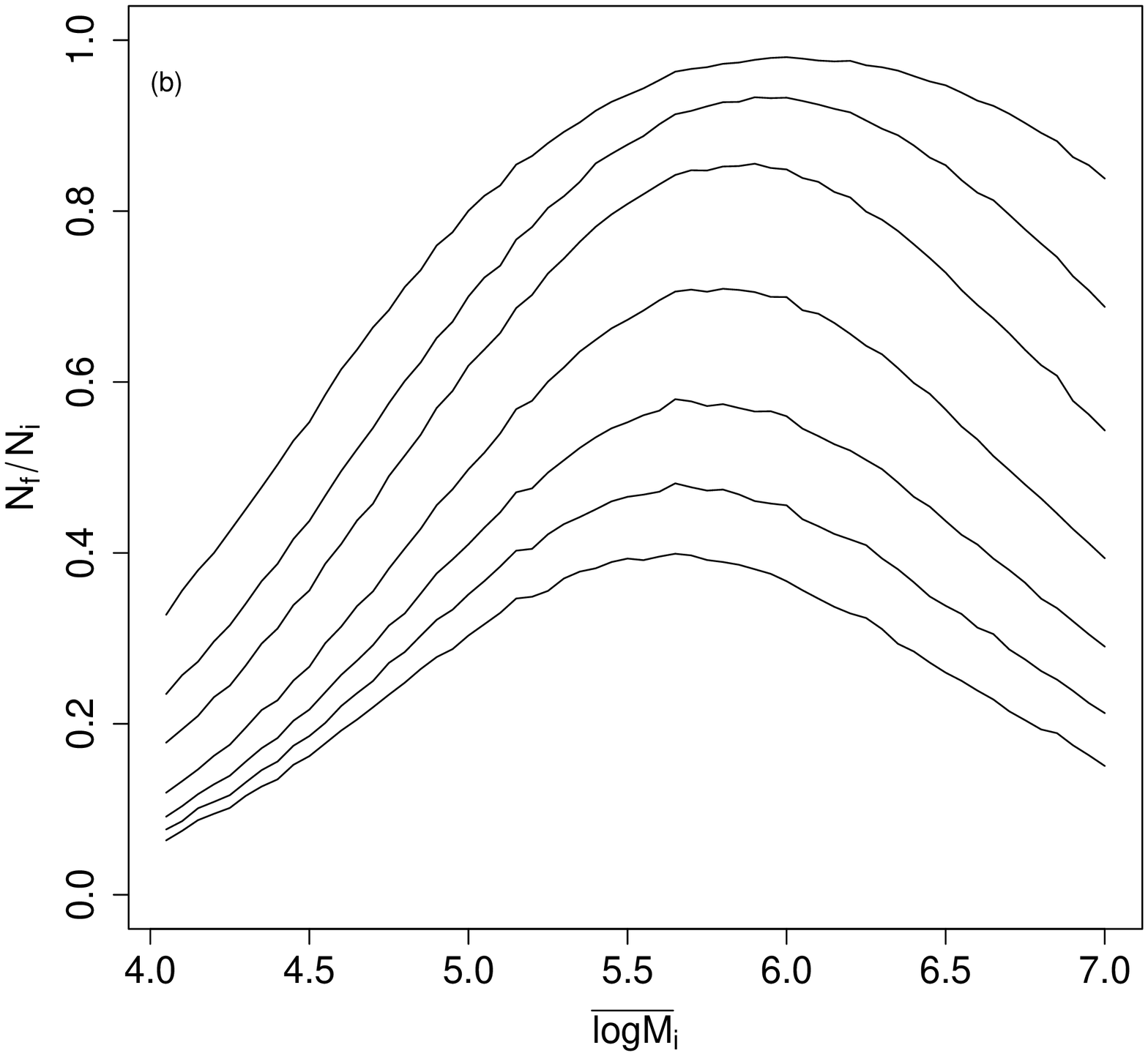,height=7cm,width=7cm,angle=0}}
\centerline{\psfig{file=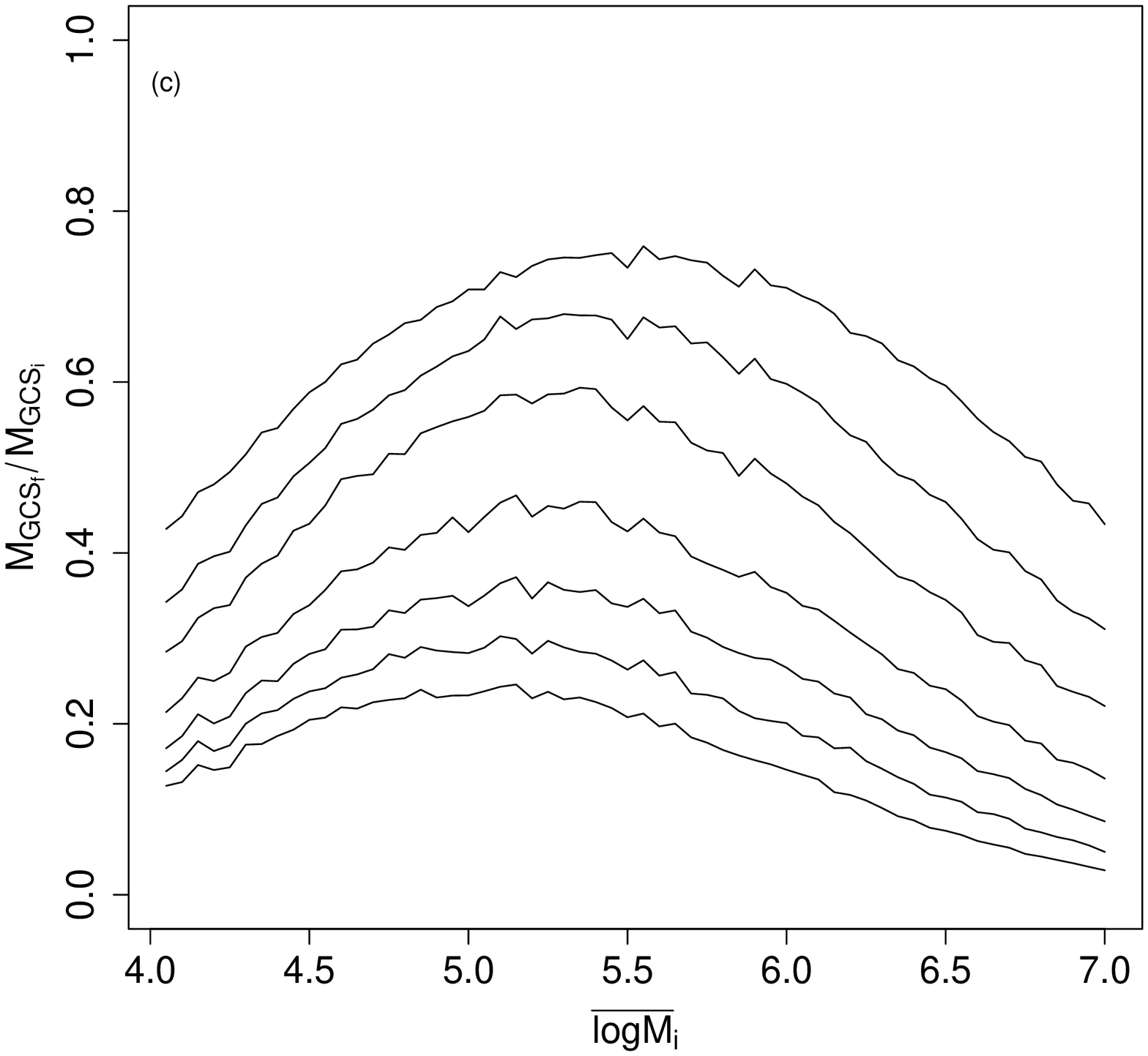,height=7cm,width=7cm,angle=0}}
\caption{(a) $\lmf$, (b) $\nfni$  and (c) $\mfmi$ versus the initial
value of $\lm$ (assuming a log-normal initial GCMF with $\sigma=0.6$) for the
seven host galaxies with $\log M_e>10.5$ discussed in \S 4. In each
panel, the seven curves shown correspond, from the upper to the lower one, to
host galaxies with decreasing values of $M_e$ (in panel (a) the curves
cross each other, and we refer to the order of the seven curves on the
right side of the plot). }
\end{figure}

As shown in Figs
13b and 13c, low-mass galaxies are generally those where  disruption processes
are more efficient and where the difference between
$\lmi$  and $\lmf$ is larger. The dependence of $\lmf$ on 
$M_{e}$ varies with $\lmi$: for low values of $\lmi$, $\lmf$ decreases as
$M_{e}$ increases while $\lmf$ increases with $M_{e}$   for high values of
$\lmi$. Since here we have restricted our attention to galaxies
with
$\log M_{e}>10.5$, for which  observational analyses show that $\lm$
spans a narrow range of values,  our
results seem to exclude the possibility that the initial GCMF was a log-normal
function with low ($\lmi\ltorder 4.7$) or high ($\lmi\gtorder 5.5$)
values of the mean mass. Moreover the values of $\lmf$ obtained for
$\lmi \ltorder 4.7$ and for $\lmi \gtorder 5.5$ do not fall in the range of
observed values.

\section{Summary and conclusions}
In this paper we have studied the evolution of globular cluster systems in
elliptical galaxies.  
We have followed the evolution of globular cluster systems  with a log-normal
initial GCMF with initial mean mass and dispersion similar to those
currently observed in the external regions of some 
galaxies where evolutionary effects are unlikely to have significantly
altered the initial
conditions ($\lmi=5.25$ and $\sg_i=0.6$ corresponding, for $M/L_V=2$,
to $M_V=-7.5$ and $\sg_V=1.5$); a large set of N-body simulations
carried out by Vesperini \& 
Heggie (1997) have been used to determine the evolution of the masses of
individual globular clusters.
\begin{itemize}
\item In the first part of the paper we have carried out a  survey
over a large sample of values of the effective mass, $M_e$, and radius, $R_e$,
of the host galaxy and we have investigated in detail the dependence
of the final 
mean mass and dispersion of the GCMF, of 
the fraction of surviving clusters and of the ratio of the final to
the initial total mass of clusters on  $M_e$ and $R_e$. The 
difference between the properties of the GCMF of inner and outer clusters has
been investigated too. Contour plots of $\lmf$, $\sgf$, $\nfni$, $\mfmi$,
$\dmio$ in the plane
$\log M_e-\log R_e$ (see Figs 2, 3 and 4) have allowed us to determine the
relation between the evolution of the GCMF and the fraction of
clusters disrupted and to show the dependence of the evolution of a
GCS on the structure of the host galaxy.
\item In the second part of the paper, we have focussed  our attention on
a subset of values of $M_e$ and
$R_e$ equal to the observational values available for giant, normal and
dwarf ellipticals and we have compared our theoretical results with the results
of some recent observational analyses (see Figs 5-10).
\begin{enumerate}
\item For galaxies with $\log M_e \gtorder 10.5$, our results are in general
qualitative and 
quantitative agreement with the findings of several observational studies:
$\lmf$ is approximately constant ($\lmf\simeq 5.16$ or $M_V=-7.3$ for
$M/L_V=2$) with a small
galaxy-to-galaxy dispersion. 
 We have shown  that the narrow range spanned by 
$\lmf$  for galaxies with different structures and the lack of a strong radial
gradient in the properties of GCMF inside many individual galaxies do {\it not}
imply that evolutionary processes have been unimportant: in 
most of the host galaxies we have considered, a significant fraction of
clusters have been disrupted and the masses of many of
those which survived have changed; in contrast with the narrow range
of values of $\lmf$, the fraction of surviving clusters for different
host galaxies, $\nfni$, spans almost the entire range of possible values.  
\item As to dwarf galaxies, our results show that evolutionary processes lead
to values of $\lmf$ smaller than those of giant ellipticals and to a
larger galaxy-to-galaxy 
dispersion. Both these findings are in agreement with those of 
observational analyses (Harris 2000).
\item The fraction of surviving clusters increases with the mass of
the host galaxy and it ranges from $\nfni\sim 0.9$ for the most
massive galaxies to $\nfni \sim 0.1$ for dwarf galaxies.  A  mass-(or
luminosity-)specific frequency correlation can result from the effects
of evolutionary processes (see also Murali \& Weinberg 1997a). 
The ratio of the final to the initial total mass of clusters ranges from
$\mfmi \sim 0.8$ for the most massive galaxies to $\mfmi \sim 0.01$ for dwarf
galaxies. 
\item The time evolution of the distribution of $\lm$, $N(t)/N_i$,
$M_{GCS}(t)/M_{GCS,i}$ for all the galaxies considered as well as the
detailed time evolution of the GCMF parameters, of $N(t)/N_i$, 
and of $M_{GCS}(t)/M_{GCS,i}$ for four fiducial galaxies have been
investigated (see Figs 8, 10, 11); this allows to determine the
expected properties of GCS in elliptical galaxies younger than the
adopted age of 15 Gyr. Our analysis shows that while the number of
clusters and their total mass, $N(t)/N_i$ and
$M_{GCS}(t)/M_{GCS,i}$, continue to decrease for 15 Gyr, the mean mass of
clusters, $\lm$, evolves significantly only in the first few Gyrs.

\item For some fiducial galaxies we have studied in larger detail the
dependence of the properties of the GCMF and of the fraction of surviving
clusters on the galactocentric distance (see Fig. 12). We have shown that
in massive galaxies 
disruption occurs mainly within
$R_e$ while in low-mass compact galaxies a significant fraction of clusters
are disrupted also at $R_g>R_e$. The difference in concentration between the
spatial distribution of clusters and that of stars is, at least in part, due to
the disruption of inner clusters.

As discussed above, we have shown that there are several systems in which the
radial gradient of
$\lmf$ is negligible in spite of a significant disruption of clusters.
\item For a sample of fiducial galaxies with $\log M_e >10.5$, we have
studied the dependence of the final properties of the GCMF  on the
initial value 
of $\lm$, $\lmi$. We have shown how the dependence of $\lmf$ on the
mass of the  
host galaxy, the spread of values of $\lmf$ and  the fraction
of surviving clusters vary with $\lmi$ (see Fig. 13). 
$\lmf$ increases as $\lmi$ increases; for a given value of $\lmi$,
the spread of values of $\lmf$ is larger for $\lmi\ltorder 4.7$
and $\lmi\gtorder 5.5$ while for intermediate values of $\lmi$
($4.7\ltorder \lmi \ltorder 5.5$) the range of $\lmf$ is small.

A small spread of values
of $\lmf$ in different galaxies is never the result of negligible disruption
or of a similar fraction of surviving clusters in the host galaxies considered.
The properties of the final GCMF are always
determined by the interplay between disruption of clusters and evolution of the
masses of the clusters which survive.
\end{enumerate}
\end{itemize}

In this paper we have focussed our attention on GCS with a log-normal initial
GCMF. In a companion paper (Vesperini 2000) we will study the
evolution of GCS starting with a power-law initial GCMF similar to
that observed in young cluster systems in merging galaxies and we will
investigate whether the final properties of GCS starting 
with such a functional form for the initial GCMF are consistent with
those currently observed in old globular cluster systems.
\section*{Acknowledgments}
I wish to thank Giuseppe Bertin and the referee, Oleg Gnedin, for many
useful comments on the paper.\\
Support form a Five College Astronomy Department fellowship is acknowledged.
\section*{References}
Aguilar L. , Hut P.  Ostriker J.P. 1988, ApJ, 335, 720\\
Ashman K.M., Zepf S.E., 1998, Globular Cluster Systems, Cambridge
University Press \\
Baumgardt H., 1998, A\&A, 330, 480\\
Bellazzini M., Vesperini E., Ferraro F.R., Fusi Pecci F., 1996, MNRAS,
279, 337\\ 
Binney J., Tremaine S., 1987, Galactic Dynamics, Princeton University
Press, Princeton, New Jersey\\ 
Blakeslee J.P., 1999, AJ, 118, 1506\\
Burstein D. , Bender R., Faber S., Nolthenius R., 1997, AJ, 114, 1365\\
Caputo F., Castellani V., 1984, MNRAS, 207,185\\
Capuzzo Dolcetta R., Tesseri A., 1997, MNRAS, 292, 808\\
Chernoff D.F., Djorgovski S.G., 1989, ApJ, 339, 904\\
Chernoff D.F., Kochanek C.S., Shapiro S.L., 1986, ApJ, 309, 183\\
Chernoff D.F., Shapiro S.L., 1987, ApJ, 322, 113 \\
Chernoff D.F., Weinberg M.D., 1990, ApJ, 351, 121\\
Crampton D., Cowley A.P., Schade D., Chayer P., 1985, ApJ, 288, 494\\
Djorgovski S.G., Santiago B.X., 1992, ApJ, 391, L85 \\
Djorgovski S.G., Meylan G., 1994, AJ, 108, 1292\\
Durrel P. Geisler D., Harris W.E., Pudritz R., 1996, AJ, 112, 972\\ 
Elmegreen, B. G., 2000, in "Toward a New Millennium in Galaxy
Morphology" D.L. Block, I. Puerari, A. Stockton and D. Ferreira editors
(Kluwer, Dordrecht), in press \\ 
Faber S.M., Dressler A., Davies R., Burstein D., Lynden-Bell D., 1987, 
in Nearly Normal Galaxies, ed. S.M.Faber (New York Springer) 175\\
Fall S.M., Rees M.J., 1977, MNRAS, 181, 37P\\
Fall S.M., Malkan M.A., 1978, MNRAS, 185, 899\\
Ferrarese L., et al., 2000, ApJ, in press\\
Forbes D.A.,Franx M., Illingworth G.D., Carollo C.M., 1996, ApJ, 467,
126\\
Forbes D.A., Brodie J.P., Hucra J., 1996, AJ, 112, 2448\\
Forbes D.A., Brodie J.P., Hucra J., 1997, AJ, 113, 887\\
Gnedin O.Y., 1997, ApJ, 487, 663\\
Gnedin O.Y., Ostriker J.P., 1997, ApJ, 474, 223\\
Gnedin O.Y., Hernquist L., Ostriker J.P., 1999, ApJ, 514, 109\\
Harris W.E., 1991, ARA\&A, 29, 543\\
Harris W.E., 2000,in  Lectures for 1998 Saas-Fee Advanced School on Star
Clusters, in press\\
Harris W.E., van den Bergh S., 1981, AJ, 86, 1627\\
Harris W.E., Harris G.L.H., McLaughlin D.E., 1998, AJ, 115, 1801\\
Jacoby G.H., et al. 1992, PASP, 104, 599\\
Kavelaars J.J., Hanes, D.A., 1997, MNRAS, 285, L31\\
Kissler-Patig M., 1997, A\&A, 319,83\\
Kundu A., Whitmore B.C., Sparks W.B., Macchetto F.D., Zepf S.E.,
Ashman K.M., 1999, ApJ 513, 733\\ 
McLaughlin D.E., 1999, AJ, 117, 2398\\
McLaughlin D.E., Harris W.E., Hanes D.A., 1994, ApJ, 422, 486\\
Meylan G., Heggie D.C., 1997, A\&A Rev., 8, 1\\
Miller B.W., Lotz J.M., Ferguson H.C., Stiavelli M., Whitmore B.C,
1998, ApJ, 508, L133\\
Murali C., Weinberg M.D., 1997a, MNRAS, 288, 767\\
Murali C., Weinberg M.D., 1997b, MNRAS, 291, 717\\
Okazaki, T., Tosa, M., 1995, MNRAS, 274, 48\\
Ostriker J.P., Gnedin O.Y., 1997, ApJ, 487, 667\\
Santiago B.X., Djorgovski S.G., 1993, MNRAS, 261, 735 \\
van den Bergh S., 1995a, AJ, 110, 1171\\
van den Bergh S., 1995b, AJ, 110, 2700\\
van den Bergh S., 1996, AJ, 112, 2634\\
Vesperini E., 1994, Ph.D. Thesis, Scuola Normale Superiore, Pisa\\
Vesperini E., 1997, MNRAS, 287, 915\\
Vesperini E., 1998, MNRAS, 299, 1019\\
Vesperini E., 2000, submitted to MNRAS\\
Vesperini E., Heggie D.C., 1997, MNRAS, 289, 898\\
Weinberg M.D., 1994a, AJ, 108,1398\\
Weinberg M.D., 1994b, AJ, 108, 1403\\
Weinberg M.D., 1994c, AJ, 108, 1414\\
Whitmore B.C., 1997, in The Extragalactic distance scale, M.Livio,
M.Donahue, N.Panagia editors, Cambridge University Press\\ 
Zepf S.E., Ashman K.M., 1993, MNRAS, 264, 611\\
Zepf S.E., Geisler D., Ashman K.M., 1994, 435, L117\\

\end{document}